\documentclass[aps,prb,reprint,groupedaddress,showpacs]{revtex4-1}
\usepackage{graphicx}  % needed for figures
\usepackage{dcolumn}   % needed for some tables
\usepackage{bm}        % for math
\usepackage{amssymb}   % for math
\usepackage{amsmath} %for math
\usepackage{color}

\begin{document}

%\preprint{}

\title{ Breakdown of Carr-Purcell Meiboom-Gill spin echoes in inhomogeneous fields}


\author{Nanette N. Jarenwattananon}
\author{Louis-S. Bouchard}
\email[]{bouchard@chem.ucla.edu}
%\thanks{}
\affiliation{Department of Chemistry and Biochemistry, University of California Los Angeles, 607 Charles E Young Drive East, Los Angeles, CA 90095-1059}


\date{\today}

\begin{abstract}
The Carr-Purcell Meiboom-Gill (CPMG) experiment has been used for decades to measure nuclear-spin transverse ($T_2$) relaxation times. In the presence of magnetic-field inhomogeneities, the limit of short interpulse spacings yields the intrinsic $T_2$ time. Here we show that the signal decay in such experiments exhibits fundamentally different behaviors between liquids and gases. In gases, CPMG unexpectedly fails to eliminate the inhomogeneous broadening due to the non-Fickian nature of the motional averaging.
\end{abstract}

%kinetic theory and transport of gases; Viscosity, diffusion, and thermal conductivity; self-diffusion (in liquids); diffusion (nmr), 
%\keywords{nuclear magnetic resonance, diffusion, motional averaging}

\maketitle

%\section{}

Previously, we found that the decay of nuclear induction signal in a magnetic field gradient differs fundamentally in gases compared to liquids, as manifested in the temperature dependence of the nuclear magnetic resonance (NMR) linewidth~\cite{jarenwattananon2015,jarenwattananon2016}.  In the conventional description of NMR, spectral lines should broaden in a gradient of the magnetic field as temperature increases~\cite{slichter1990principles}, which results in a larger diffusion coefficient. However, we found experimentally that in gases, the NMR linewidth instead {\it decreases} with temperature, which is consistent with a motional narrowing effect.  The importance of this motional narrowing effect was not predicted by the conventional theory.  In this article, we demonstrate that this difference in motional averaging between gases and liquids also manifests itself in the signal decay of CPMG spin echoes. In liquids, a series of spin echoes in the limit of short interpulse spacings minimizes signal decay effects due to diffusion in a gradient (as expected from the conventional theory~\cite{slichter1990principles}).  For gases, however, we find that CPMG is unable to eliminate this signal decay in the limit of short interpulse spacings.  This result  implies that any  $T_2-$  or diffusion-weighted NMR measurements of gases made in the presence of magnetic susceptibility gradients or applied field-gradients are potentially compromised.

\begin{figure}[h!]
\includegraphics[width=4cm]{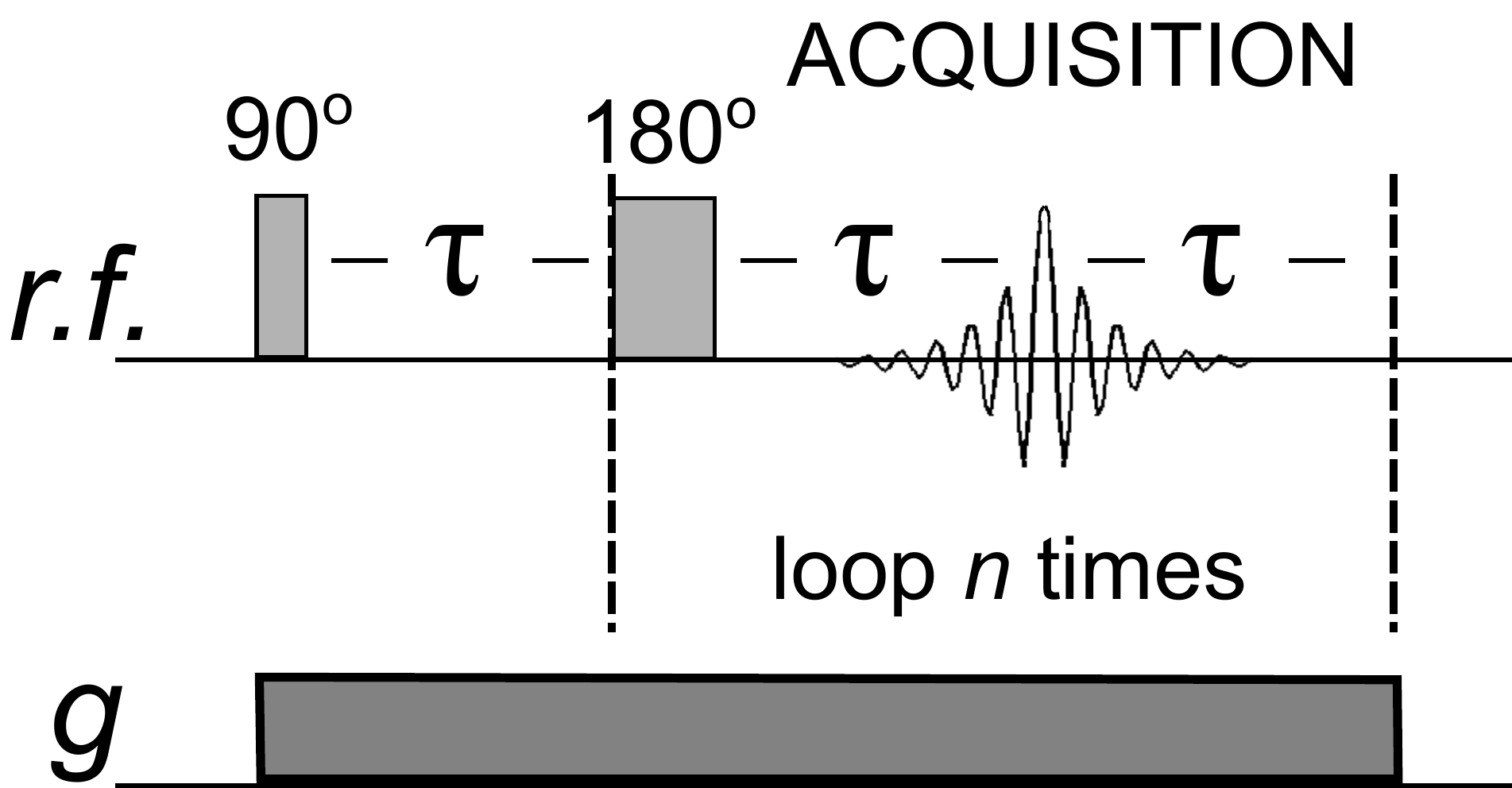}
\caption{\label{fig:pulseseq}Measurement of $T_2$ in an inhomogeneous field (gradient, \textit{g}) using a CPMG sequence.  A $90^{\circ}$ radio-frequency (r.f.) pulse tips the magnetization, which is refocussed by a series of \textit{n} $180^{\circ}$ pulses at intervals of $2\tau$.}
\end{figure}

In a 90$^\circ-\tau-$180$^\circ-\tau$ spin-echo experiment the free induction signal in the presence of a magnetic field gradient is given by Hahn's famous result~\cite{hahn1950,slichter1990principles}:
\begin{equation}
S(\tau) = \exp{(-2\tau/T_2)}\exp{\bigg(-\frac{2}{3}\gamma^2g^2D \tau^3\bigg)},
\label{eq:signaldecay}
\end{equation}
\noindent in which $2\tau$ is the evolution time (total length of the spin-echo experiment), $T_2$ is the intrinsic spin-spin relaxation time, $\gamma$ is the nuclear gyromagnetic ratio, $D$ is the self-diffusion coefficient, and $g$ is the magnetic field gradient ($g=\partial H_z/\partial r$, where $r$ is a spatial direction).   In the absence of a large magnetic field inhomogeneity and large diffusivity the expression collapses to $\exp{(-2\tau/T_2)}$.   The result~(\ref{eq:signaldecay}) assumes that the Einstein-Fick limit holds, implying that we can make the approximation to the position autocorrelation function (PAF), $\langle x(t)x(0)\rangle \approx \langle x(t) x(t)\rangle =2Dt$, i.e., the PAF is approximated by the mean-square displacement. 

According to the same conventional theory, the signal in the CPMG experiment, which features a train of $n$ echoes~(Fig.~\ref{fig:pulseseq}), decays according to ~\cite{slichter1990principles,carr1954,meiboom1958}:
\begin{equation}
S(\tau) = \exp{(-n2\tau/T_2)}\exp{\bigg(-\frac{1}{3}\gamma^2g^2D \tau^2 (n2\tau)\bigg)}, 
\label{eq:cpmg}
\end{equation}
\noindent where $n2\tau$ is the total duration of the sequence, which is held fixed.    In practice, an upper bound on $2n\tau$ is imposed by the relaxation time of the sample.   In the limit of short echo spacing ($\tau \rightarrow 0$, holding $n2\tau$ constant) CPMG minimizes the effect of molecular self-diffusion on nuclear spin decoherence in inhomogeneous magnetic fields. Under these conditions, the contribution of the inhomogeneous term becomes negligible, recovering $\exp{(-n2\tau/T_2)}$.

However, the expression~(\ref{eq:cpmg}) only holds for substances whose diffusional properties obey the Einstein-Fick limit, which mainly applies to liquids. Gases typically lie outside the Einstein-Fick limit.  We have derived in prior work an expression for signal decay in gases~\cite{jarenwattananon2015,jarenwattananon2016}:
\begin{equation}
S(\tau) = \exp{(-2\tau/T_2)}\exp{\bigg(-\gamma^2g^2\kappa\tau\bigg)}, 
\label{eq:gasdecay}
\end{equation}
\noindent where $\kappa$ is a term that depends on the PAF of  diffusing molecules.  It depends, among other things, on temperature and viscosity.  We note that the time dependence is $\tau$, not $\tau^3$.   This difference in exponents has important consequences.  Namely, the application of a CPMG sequence with $n$ echoes,
\begin{equation}
S(\tau) = \exp{(-2n\tau/T_2)}\exp{\bigg(-\gamma^2g^2\kappa(2n\tau)\bigg)}. \label{eq:gascpmg}
\end{equation}
\noindent no longer eliminates the second term describing inhomogeneous-field decay in the limit of short interpulse spacing $\tau \rightarrow 0$ ($2n\tau$ is fixed).   Thus, any measurement of  $T_2$ in a gas using a CPMG sequence will yield an {\it apparent} $T_2$ value that is affected by diffusivity effects in the inhomogeneous magnetic field.   This could include, for example, unwanted weightings due to temperature, viscosity, external hardware, pulse sequence design, magnetic susceptibility and pore geometry.  This is in contrast to the case of liquids, where  diffusion effects can be mitigated by extrapolation to extract the true (intrinsic) $T_2$ time.

Consider the CPMG experiment~(Fig.~\ref{fig:pulseseq}) with a $90^{\circ}$ broadband pulse and a series of $180^{\circ}$ pulses to refocus the magnetization at intervals of $2\tau$.  The following phases were applied in the CPMG sequence: ${90^\circ}_x -  \tau - ({180^\circ}_y - \tau - \tau)^n$, where the $180^\circ$ pulse is repeated $n$ times. For each $\tau$ value, the echo envelope was acquired in a single-shot experiment. An external magnetic field gradient during the course of the experiment creates an inhomogeneous magnetic field.  For gas-phase experiments, a sealable 5-mm diameter J. Young NMR tube was filled with liquid and freeze-pump-thawed to evacuate excess air.  The NMR tube was heated by the NMR spectrometer's variable temperature unit until the tube was vaporized. For liquid-phase experiments, a solution of 0.5$\%$ weight/volume tetramethylsilane (TMS) in acetone-$d_6$ was degassed and flame-sealed in an NMR tube. Measurements were performed on a 14.1~T vertical bore Bruker AV 600 MHz NMR spectrometer equipped with a 5 mm broadband probe with a $z$-gradient. The receiver was operated in qsim mode (forward Fourier transform, quadrature detection).  The pulses were hard pulses whose lengths were 16~$\mu$s and 32~$\mu$s for the 90$^\circ$ and 180$^\circ$ pulses, respectively.  The size of the sensitive RF region is less than 1 cm.     All NMR signals were analyzed in magnitude mode and decay functions included baseline subtraction.

\begin{figure}
\includegraphics[width=\columnwidth]{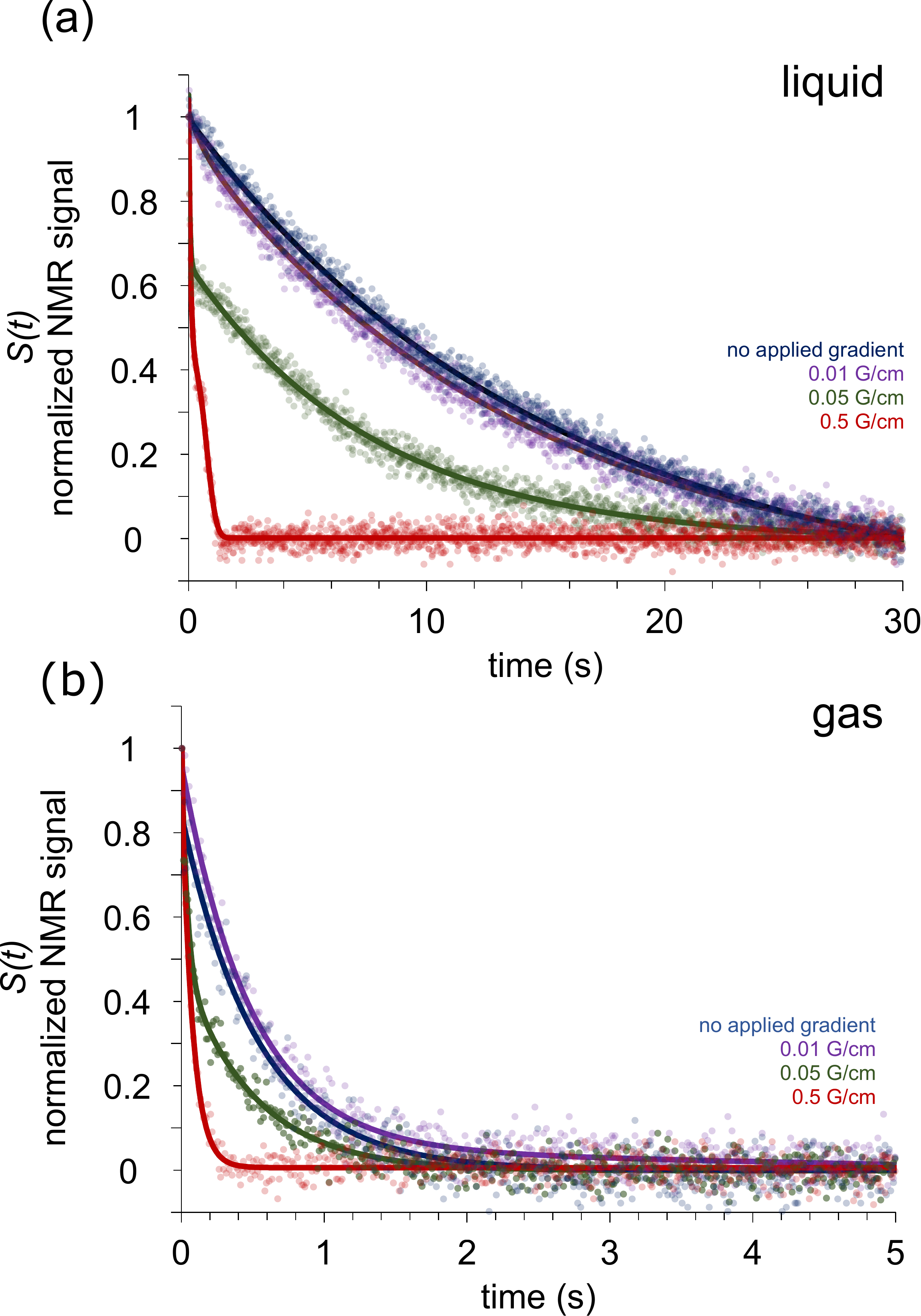}
\caption{\label{fig:gasliquid} 
Direct verification that signal decay in the presence of a magnetic-field gradient follows a time dependence of the form $\exp(-t/T_2)\exp(-t/b)$ for the gas and $\exp(-t/T_2)\exp(-(t/b_2)^3)$ for the liquid (solid lines, fit; dots, data).  Here we show sample CPMG decay curves for $\tau$=5 ms and $g$=0, 0.01, 0.05 and 0.5 G/cm.  (a) Liquid-phase TMS ($t^1$). Number of scans (ns) = 1. (b) Gas-phase TMS ($t^3$).  ns = 8.  In both liquid and gas cases, we scanned multiple acquisitions (ns = 1 and 8 for liquid and gas, respectively) and a $T_2$ value with experimental error bars was derived for Figure 3. The fits to the respective models are excellent.  Fits to the converse equation ($t^1 \leftrightarrow t^3$) do not yield acceptable fits (not shown here).}
\end{figure}

We take an explicit look at the time-decay of the CPMG echo train, which the theory (c.f. Eq. \ref{eq:signaldecay} and \ref{eq:gasdecay}) predicts should exhibit fundamentally different behavior ($t^3$ vs $t^1$, respectively).  A direct verification is obtained by plotting the NMR signal in the CPMG experiment versus time along the echo train (see Fig.~\ref{fig:gasliquid}). The normalized NMR signal decay of the CPMG spin echo (with an interpulse spacing of 5 ms) for liquid-phase TMS is plotted in Fig.~\ref{fig:gasliquid}a. The normalized NMR signal decay of the CPMG spin echo (with an interpulse spacing of 5 ms) is plotted in Fig.~\ref{fig:gasliquid}b for gas-phase TMS. For gas, the normalized NMR signal decays exponentially, according to $\exp(-t/T_2) \exp(-t/b)$, in agreement with our revised theory of the NMR linewidth.   We note that neither alteration of the phase cycling scheme to ${90^\circ}_x - ({180^\circ}_y -  {180^\circ}_{-y})_n$ nor replacement of the ${180^\circ}$ pulse with a ${90_x^\circ} / {180_y^\circ} / {90_x^\circ}$ composite pulse affected the results.

\begin{figure*}
\includegraphics[width=6in]{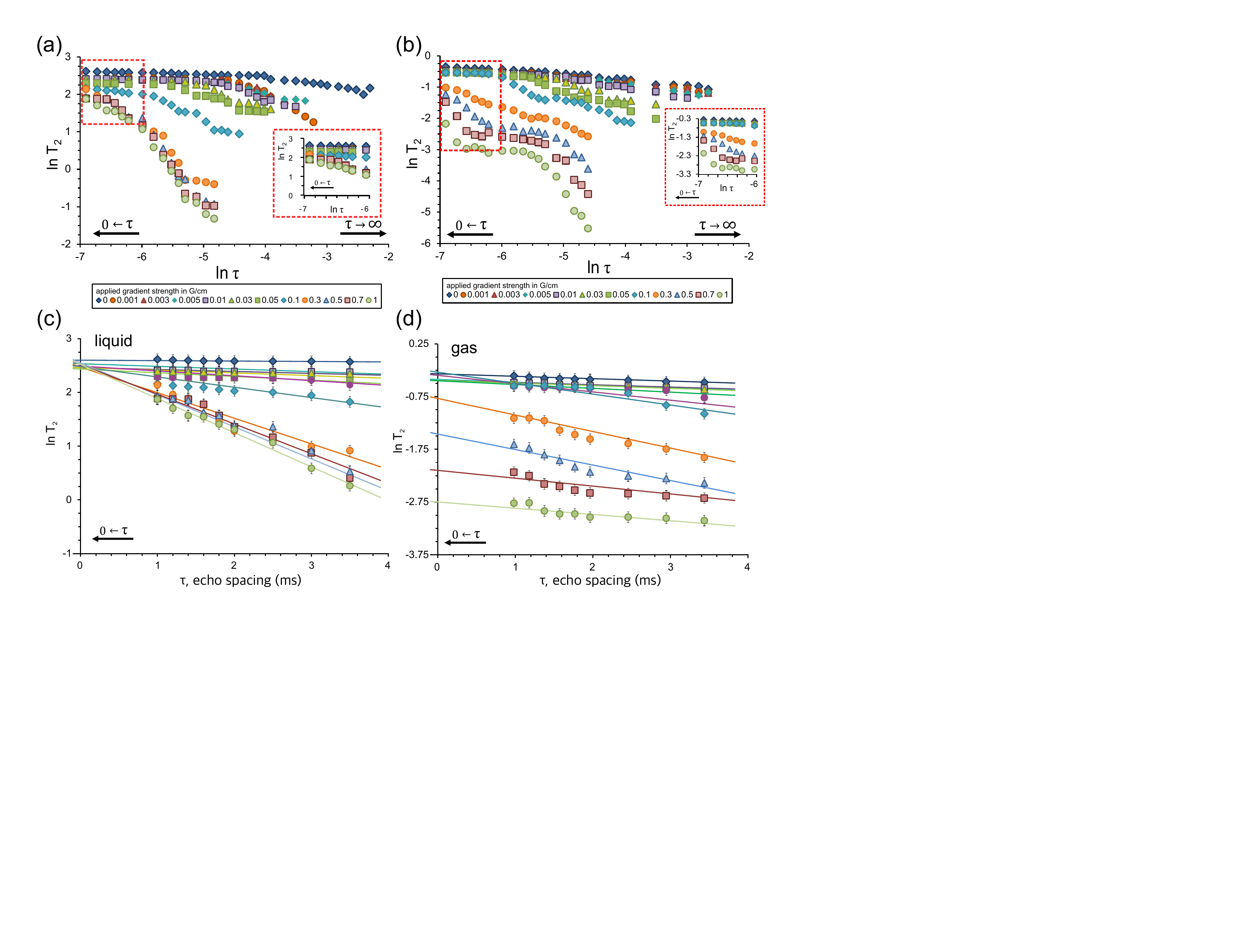}
\caption{\label{fig:combined} $T_2$ relaxation time of tetramethylsilane (TMS) vs interpulse spacing $\tau$ under conditions of magnetic field inhomogeneity (field gradient $g$, in G/cm). (a) Liquid-phase $T_2$ values approach the limit of no applied gradient as $\tau \rightarrow 0$. $T_2$ values shown range from 50 ms to 20 s. $\tau$ values shown range from 1 ms to 100 ms. Inset: Expansion of red boxed region.
(b) Gas-phase $T_2$ values do not converge to a single value as $\tau \rightarrow 0$. $T_2$ values shown range from 18 ms to 1 s. $\tau$ values shown range from 1 ms to 100 ms. Inset: Expansion of red boxed region in A. 
(c) Data from (a) plotted on a linear scale.  The straight lines are linear extrapolations as $\tau \rightarrow 0$. The $g$ values are the same as in (a).
(d) Data from (b) plotted on a linear scale.  The straight lines are linear extrapolations as $\tau \rightarrow 0$. The $g$ values are the same as in (b). }
\end{figure*}

Measurements of the CPMG echo train signal decay as a function of interpulse spacing $\tau$ for TMS in the liquid phase are shown in~Fig.~\ref{fig:combined}a.   Figure~\ref{fig:combined}b shows the corresponding experiment in the gas phase.  TMS is a liquid at room temperature but a gas at 26$^\circ$ C; thus, a modest temperature increase enables comparison of the same substance in two different phases.  TMS  was also chosen due to its long relaxation times in both liquid and gas phases, enabling us to apply a large number of refocusing pulses even at long $\tau$ values.  For liquids, as the interpulse spacing decreases the measured $T_2$ value approaches a single value irrespective of the applied gradient strength $g$, as if there were no external gradient~(Fig.~\ref{fig:combined}a,c).   This corresponds to the limit $\tau \rightarrow 0$ in Eq.~\ref{eq:cpmg}.  Extrapolation of $T_2$ to the limit $\tau \rightarrow 0$ is the most commonly used method to extract true $T_2$ times in the presence of magnetic-field inhomogeneities (from external or internal fields). For gases, however, the $T_2$ values in the limit $\tau\rightarrow 0$ do not approach a single value (Fig.~\ref{fig:combined}b,d), but instead converge to different values depending on the applied gradient strength $g$.  This fundamentally different behavior implies that the inhomogeneous-field decay term is still present, as predicted by Eq.~\ref{eq:gascpmg}.

%In liquids, a second exponential decay term can be observed for certain choices of interpulse spacing and gradient amplitude, so that the decay is of the form $A \exp(-t/b)+B\exp(-(t/b_2)^3)$.   The first term, $A \exp(-t/b)$, arises from frequency-encoding effects~\cite{jarenwattananon2015}.   Frequency-encoding from a gradient of amplitude $g$ and duration $t$ results in a sinusoidal modulation of the magnetization at wavelength $\lambda=2\pi/(\gamma g t)$.  In our experiments $\lambda$ was as low as 34 $\mu$m at the strongest gradient strength (1 G/cm) and longest interpulse spacing (70 ms).  For liquid TMS the rms diffusion length ranged from 3 $\mu$m ($\tau$=1 ms) up to 25 $\mu$m ($\tau$=70 ms).   For gas-phase TMS the rms diffusion length ranged from 20 $\mu$m ($\tau$=1 ms) up to 160 $\mu$m ($\tau$=70 ms).  The longer rms diffusion lengths in the gas phase result in a wiping out of the magnetization modulation and no frequency-encoding effect is observed.   In  liquid phase TMS,  the magnetization modulation is only partially removed by diffusion and  a frequency-encoding term describes the decay.

In this study we have confirmed the $t^1$ dependence in the NMR signal decay function of gases in the presence of an external gradient (Eq.~\ref{eq:gasdecay} and ~\ref{eq:gascpmg}).  In our prior work we had verified the temperature dependence of the linewidth~\cite{jarenwattananon2015}.  The verification of the $t^1$ time dependence can be considered the missing part of the puzzle which now unambiguously confirms the validity of the revised linewidth theory presented in Ref.~\cite{jarenwattananon2015}.  The $g^2$ dependence has already been verified in our previous paper~\cite{jarenwattananon2015}.

The fundamentally different motional averaging behavior of the NMR experiment in gases has important implications for several experiments.   This behavior has previously led to the development of a novel non-invasive method for mapping temperatures of gases~\cite{jarenwattananon2013}.   Gas-phase MRI experiments that utilize frequency-encoding gradients could be affected; gradients during readout affect measurements of $T_2$ or diffusion, introducing an apparent coupling between them.  This means that a quantitative interpretation of these parameters would need to account for the non-Fickian nature of the diffusion. Dynamic decoupling schemes such as the Uhrig sequence~\cite{uhrig2007}, which aim at generating the longest possible coherence times, are also expected to break down in the case of gases because short $\tau$ values no longer guarantee the elimination of environmental factors.   Finally, the interpretation of restricted diffusion results~\cite{callaghan2011,grebenkov2007nmr, Zhang200381,zielinski2005probing, mair1999probing, hurlimann1995spin, de1994decay, sen1999spin, PhysRevB.46.3465} in porous media and other confined geometries may require new theoretical developments that model the signal decay in restricted environments in light of the new theory of signal decay.

%\section{Methods}
%{\it Methods:} 

%\begin{figure}
%\includegraphics[width=4cm]{fig1.pdf}
%\caption{\label{label} Caption}
%\end{figure}

\begin{acknowledgments}
This work was partially funded by a Beckman Young Investigator Award and the National Science Foundation through grants CHE-1153159 and CHE-1508707.
\end{acknowledgments}	

\bibliography{references}

\end{document}